Discovery of Lake-effect clouds on Titan

M.E. Brown[1], E.L. Schaller[1], H.G. Roe[2], C. Chen[1,3], J. Roberts[1], R.H. Brown[4], K.H. Baines[5], R.N. Clark[6]

**Abstract:**

Images from instruments on Cassini as well as from telescopes on the ground reveal the presence of sporadic small-scale cloud activity in the cold late-winter north polar of Saturn's large moon Titan. These clouds lie underneath the previously discovered uniform polar cloud attributed to a quiescent ethane cloud at ~40 km and appear confined to the same latitudes as those of the largest known hydrocarbon lakes at the north pole of Titan. The physical properties of these clouds suggest that they are due to methane convection and condensation. Such convection has not been predicted for the cold winter pole, but can be caused by a process in many ways analogous to terrestrial lake-effect clouds. The lakes on Titan are a key connection between the surface and the meteorological cycle.

[1] Division of Geological and Planetary Sciences, California Institute of Technology, Pasadena, CA 91125 USA

[2] Lowell Observatory, 1400 W. Mars Hill Rd., Flagstaff, AZ 86001 USA

[3] Westridge School, 324 Madeline Dr., Pasadena, CA 91105 USA

[4] Lunar and Planetary Laboratory, University of Arizona, Tucson, AZ 85721 USA

[5] Jet Propulsion Laboratory, Pasadena, CA 91109 USA

[6] US Geological Survey, Mail Stop 964, Box 25046 Federal Center, Denver, CO 80225 USA

# 1. Introduction

Tropospheric cloud cover on Titan is significantly more sparse and more variable than on the Earth, with typical surface coverage being less than 1%, but with large storms occasionally covering up to 10% of the surface [*Griffith et al.*, 1998; *Griffith et al.*, 2000; *Schaller et al.*, 2006a]. The locations of the clouds appear to be controlled by solar insolation. During southern summer solstice, when the point of maximum insolation was the pole itself, clouds were a persistent presence at high southern latitudes [*Bouchez and Brown*, 2005; *Brown et al.*, 2002]. As southern summer has waned to southern fall, the south polar clouds have mostly dissipated [*Schaller et al.*, 2006b] and clouds at southern mid-latitudes have become more frequent [*Roe et al.*, 2005]. While the time scale for radiative heating or cooling of Titan's atmosphere is so long that seasonal effects were initially thought to be irrelevant [*Hunten et al.*, 1984], it is now seen that the general seasonal behavior of Titan's clouds is moderately well reproduced with general circulation models (GCMs) where Titan's surface has a low enough thermal inertia that the seasonally changing surface temperature controls the large scale seasonal circulation changes [*Mitchell et al.*, 2006; *Mitchell*, 2008]. Clouds appear in regions of uplift where air parcels become saturated with methane which then condenses out as they ascend through the troposphere.

In this simple picture of clouds and circulation, no clouds would be expected near the winter pole where the air is primarily descending from the stratosphere. On Titan, however, other condensable species – the most abundant being ethane – are produced in the stratosphere by photolysis of methane. If air containing these species subsides across the cold tropopause a separate type of cloud system can form [*Rannou et al.*, 2006]. Such clouds were recently observed in the late winter near the pole [*Griffith et al.*, 2006]. These clouds are distinctly different from the previously observed clouds. They are near the ~40 km height of the tropopause, rather than at ~20 km in the middle of the troposphere; they appear to be composed of small particles which do not scatter efficiently at wavelengths of 5 microns and longer, in contrast to the much larger particles found in the other clouds; and they are spatially and temporally homogenous [*Griffith et al.*, 2006]. These characteristics are precisely those expected for a cloud formed from condensation of higher order hydrocarbons as they subside across the tropopause [*Griffith et al.*, 2006].

While this north polar cloud has appeared continuously since it was first discovered, a careful examination of images from the Cassini spacecraft reveals that a separate type of cloud can also be seen sporadically at the north pole. In this study we examine images and spectra of this new cloud system and discuss their origin and implications.

# 2. Observations

We examined the north polar clouds of Titan using data from VIMS and ISS instruments on board the Cassini spacecraft and from adaptive optics observations from the Gemini observatory and full-disk spectroscopy of Titan from the NASA Infrared Telescope Facility (IRTF). The Cassini data were obtained from the Planetary Data Systems (PDS) imaging archive (www.pds-imaging.jpl.nasa.gov) and covered 36 Titan flybys from July 2004 until August 2007. The VIMS data were calibrated using routines provided by the PDS. No calibration routines appear for ISS data, so the data are used in raw form.

North polar knots and streaks are identified in the VIMS data by constructing a tropospheric image out of each hyperspectral image by summing the data from the channels at wavelengths between 2.83 and 2.90 microns, all of which are wavelengths of moderate methane opacity where photons reach the troposphere but not the surface. We examine each image and look for spatial inhomogeneities in the north polar cloud. A feature is not labeled as a north polar knot or streak unless it appears in multiple images and multiple pixels.

Clouds are identified in the ISS images by morphological appearance alone, as no information is available on their heights. Locations of clouds are determined by comparison of the surface image to a surface base map released by the ISS team at www.ciclops.org

The Gemini data were obtained in a continuation of a long term monitoring program of Roe et al. [2005] and calibrated identically to the earlier program. With the much lower spatial resolution of these ground-based images no individual north polar features are resolved, but, on occasion, the north polar region temporarily brightens dramatically in the 2.11-2.137 micron filter which is sensitive to scattering in troposphere and above. While the ground-based data are significantly less sensitive to these cloud features, the much more frequent imaging from the ground allows a better temporal understanding of the largest of these cloud outbursts.

Spectral monitoring of Titan was performed by the IRTF telescope and the SpeX instrument in the program described by Schaller [2008]. Synoptic variations in the full disk spectra were used to monitor changes in total cloud coverage and to measure the heights of clouds detected. On several occasions, simultaneous imaging from Gemini and spectra from IRTF allow us to both pinpoint the location of a cloud and measure its altitude.

## 3. North polar knots and streaks

The large ethane-like north polar cloud appears essentially unchanged in every image of the north pole of Titan obtained since its discovery, but a close look shows small bright knots or steaks within the otherwise homogeneous north polar cloud. With the small amount of time that any spot on Titan is imaged during any one flyby, little short term temporal information is available, but on one occasion images taken 5.5 hours apart show that new bright knots are capable of appearing on time scales of hours (Fig. 1). The earliest that such a knot was observed was 14 February 2005; they have since been observed in almost every Titan flyby since February 2007. The presence of these brightenings within the otherwise homogeneous north polar cloud has also been seen from ground based adaptive optics images with the Gemini telescope. While the ground-based images have lower resolution and poorer viewing geometry, the brightest north polar events can nonetheless be detected, and these have been seen on 14 occasions since 13 April 2007. On 4 of those occasions we have simultaneous full-disk spectra of Titan from the IRTF which show that the brightenings seen in the Gemini images are confined to wavelengths shortward of 2.16 microns, as has typically been seen for Titan's south polar and mid-latitude clouds [*Schaller*, 2008].

Detections of the clouds are confined to regions north of 55 N latitude, precisely the same latitudes as the largest known lakes [*Hayes et al.*, 2008]. Cassini has seen such clouds almost exclusively between longitudes of about 60 and 240 E. The Cassini detections are highly biased, however, as 10 of the 11 flybys from February 2007 until August 2007 – when most of the north polar streaks and knots have been observed – have covered the same longitude range. The 19 July 2007 flyby does cover the opposite side of Titan, however, and no knots or steaks were observed. The ground-based images are unbiased in longitude, however, and show no statistically significant longitudinal preference (Figure 2).

## 4. Spectral Analysis

Similar cloud structures were previously suggested to be simply the otherwise homogeneous north polar cloud being affected by polar winds [*Le Mouelic et al.*, 2008]. Spectral analysis suggested that the properties of the knots and streaks were indistinguishable from those of the north polar cloud [*Le Mouelic et al.*, 2008]. This analysis was, however, based on images and spectra obtained far enough away from Titan that it was difficult to spectrally distinguish the knots and streaks from the background cloud.

On 29 June 2007, however, the VIMS instrument [*Brown et al.*, 2004] on board Cassini, fortuitously obtained a high spatial resolution hyperspectral image of a north polar cloud streak, allowing us to carefully assess its spectral properties. We find that the spectrum of the streak differs significantly from the background north polar cloud. The streak is bright at 5 microns, while the north polar cloud is undetectable at these wavelengths, implying that, while the north polar cloud must be composed of particles that are smaller than 5 microns, particles that make the streak must be larger (Fig. 3).

In addition to being composed of smaller particles, the streak appears to be significantly lower in the atmosphere than the north polar cloud. The height of clouds in Titan's atmosphere can be determined by the wavelengths at which the cloud can be seen. If a cloud appears at wavelengths of strong methane opacity it must be high in the atmosphere where little methane is above it. If a cloud does not appear until wavelengths of low methane opacity it is much lower in the atmosphere. With the hyperspectral images from VIMS, the location where a cloud first appears is easily determined. We examine, in particular, the well-studied 2.0-2.23 micron region where most of the ground-based work has been done. Little methane opacity exists at 2.02 microns, so images at the wavelength see all the way to the surface. By 2.23 microns, in contrast, the methane opacity is so high that Titan is dark except for any scattering high in the stratosphere.

Between 2.02 and 2.23 microns the methane opacity increases. At 2.02 microns the surface, north polar streaks, and north polar clouds are all visible. By 2.10 microns the surface is no longer visible, but the north polar streak and north polar cloud can still be seen. The north polar steak is visible until wavelengths of 2.15 microns; by 2.17 microns only the north polar cloud can be seen. The north polar cloud remains visible until 2.20 microns, when it, too, disappears due to high methane opacity. This spectral structure demonstrates conclusively that the north polar streaks are above the surface but below the north polar cloud. Indeed, the wavelengths at which the north polar streaks are visible are identical to those at which Titan's mid-latitude convective clouds are also visible [*Griffith et al.*, 2005], demonstrating that they are at similar altitudes of 20-30 km. A similar conclusion is reached by examining the simultaneous ground-based images and spectra, which show that the northern brightening are confined to the heights as has previously been seen for south polar and mid-latitude clouds [*Schaller*, 2008].

## 5. Lake-effect clouds

The properties of these north polar streak show that they are a distinct phenomenon from the north polar cloud. In fact, these properties suggest a specific cause for this new phenomenon. The arguments put forth to suggest that the north polar cloud is caused primarily by ethane subsidence can be reversed to suggest that the north polar streaks are caused by methane convection. The larger particle size for the streaks is expected for the dominant condensable species in the atmosphere; the high spatial and temporal variability is expected for convective clouds; and the ~20-30 km altitude of the clouds is the expected height for convective clouds in Titan's troposphere [*Griffith et al.*, 2000]. No other process can readily explain all of these properties.

While the properties of these north polar knots and streaks suggest that they are convective clouds, with the low insolation at the north pole in the late winter (Titan will not reach equinox until August 2009), the air should be stably stratified and no convection should occur. Convection in otherwise stable air can be initiated, however, not just by heating, but also by an increase in humidity of an air parcel. On the Earth, winter convection is frequently seen over and downwind of lakes such as the Great Lakes in what are called lake-effect clouds. Terrestrial lake-effect clouds typically occur when cold stably stratified air passes over warmer lake water, causing an increase in air humidity and temperature, leading to condensation in an expanded convecting boundary layer. Such clouds typically have streak-like morphology resulting from secondary flows within the planetary boundary layer [*Brown*, 1980].

On Titan, lake-effect clouds will have some underlying differences from their terrestrial counterparts. Due to the small seasonal temperature differences and strong cooling effect from evaporation [*Mitri et al.*, 2007], it is unlikely that the lake temperature will ever exceed air temperatures. The lake will cool from evaporation until saturated air at the temperature of the lake is no longer buoyant compared to near-surface air, at which point turbulent exchange of heat and humidity will quickly shut-off [*Mitchell*, 2008]. For near-surface air at a temperature of 90.5

K, as predicted in GCM simulation with plausible values of surface thermal inertia for the mid-winter season before any possible lake-effect clouds were first observed [*Tokano*, 2005], the lake will cease evaporation – and possible formation of any clouds – when the surface lake temperature reaches about 90 K. In the absence of any additional energy inputs the lake would stabilize near this temperature, and evaporation – and thus the possibility of any lake-effect clouds – would cease. In the late-winter season at which these north polar knots first began to appear, however, the small increase in insolation allows evaporation to commence. In 2007 during Titan's late winter the surface at 65 N latitude on Titan was receiving approximately 0.1 W $m^{-2}$ of solar insolation, which would slowly heat the lake surface. When the lake temperature reaches the threshold at which lake-temperature air is again buoyant, significant evaporation will recommence, lowering the lake temperature back below the threshold value. This unstable system will lead to sporadic evaporation as individual lakes go through cycles of radiative heating, evaporative cooling, and turbulent shut off, as the lake surface temperature stays close to the threshold value. The evaporation will peak at summer solstice when the maximum insolation keeps the lake temperature above the turbulent threshold temperature for the maximum amount of time. The fact that extremely large north polar events have only recently begun to be observed from the ground-based Gemini images suggests that a noticeable increase in evaporation is already occurring.

Air passing over an evaporating lake will significantly increase in methane humidity. We construct a simple bulk aerodynamic model to estimate the magnitude of the humidity increase. Our model follows an air parcel as it crosses an expanse of liquid. At each time step, the mass per unit area evaporated into the parcel is calculated as

$$E = ku\rho_{air} f(q_{sat} - q)\Delta t,$$

where *E* is the evaporation rate, *k* is a dimensionless transfer coefficient describing the efficiency of turbulent exchange, $\rho_{air}$ is the density of near-surface air, *f* is the fraction of methane within a lake, *q* is the methane mole fraction, $q_{sat}$ is the methane mole fraction for a saturated parcel, and Δt is the model time step. Values for all parameters are taken from Mitri et al. [2007].

The methane is assumed to instantaneously mix within a boundary layer of height z, giving a new value of specific humidity to the parcel.

For a lake with a length of 1000 km, an air relative humidity of 50%, and a boundary layer of 300 m, and a methane mole fraction within the lake of 0.35, passing air will reach a relative humidity of 75%.

Such a parcel is in a regime where small perturbations can initiate deep convection [*Barth and Rafkin*, 2007]. With no heat sources present, the parcel would need to be lifted to ~1 km (depending on the precise yet unknown thermal structure of the bottom of the winter troposphere) before it condenses, and lifted further to ~2 km before condensation causes enough latent heat release to render the parcel buoyant and thus initiate convection [*Griffith et al.*, 2000]. A more humid parcel will condense and initiate convection even more readily. Air that circulates and crosses the large lake multiple times (or that crosses multiple lakes or saturated surface regions with equivalent total distance) will have its relative humidity increased to ~90%. Convection in this case can easily be initiated by mechanical lifting over polar topography, which appears to be of the order of ~0.5 km in the single limited swath observed with Cassini altimetry [*Hayes et al.*, 2008], but could be much larger given the large lakes and basins.

The discovery of these lake-effect clouds on Titan drives home both the similarities and differences between the hydrological cycles on Earth and on Titan. On the Earth, the formation of lake effect clouds is greatly aided by the additional buoyancy provided by the heat exchange between warm lakes and cold overlying air. On Titan, the solar insolation is so small that evaporative cooling dominates over any heating, so the lakes are never warmer than the overlying air, thus, on Titan an extra lifting mechanism – presumably mechanical lifting over topography – is required. Likewise, on Earth the global effect of lake-effect clouds is modest, as the systems generally remained confined to the boundary layer and quickly dissipate. On Titan, in contrast, the clouds break out of the boundary layer initiating deep convection at that winter pole and depositing significant heat high in the troposphere, changing the nature of Titan's circulation [*Mitchell*, 2008]. Finally, while terrestrial lake-effect clouds are predominantly a winter event, on Titan, lake-effect clouds, being controlled by evaporative flux rather than by heating from a warm lake, should increase precipitously with increased insolation. With Titan's extensive north polar lakes, northern summer should be a time of dramatic northern cloud events.

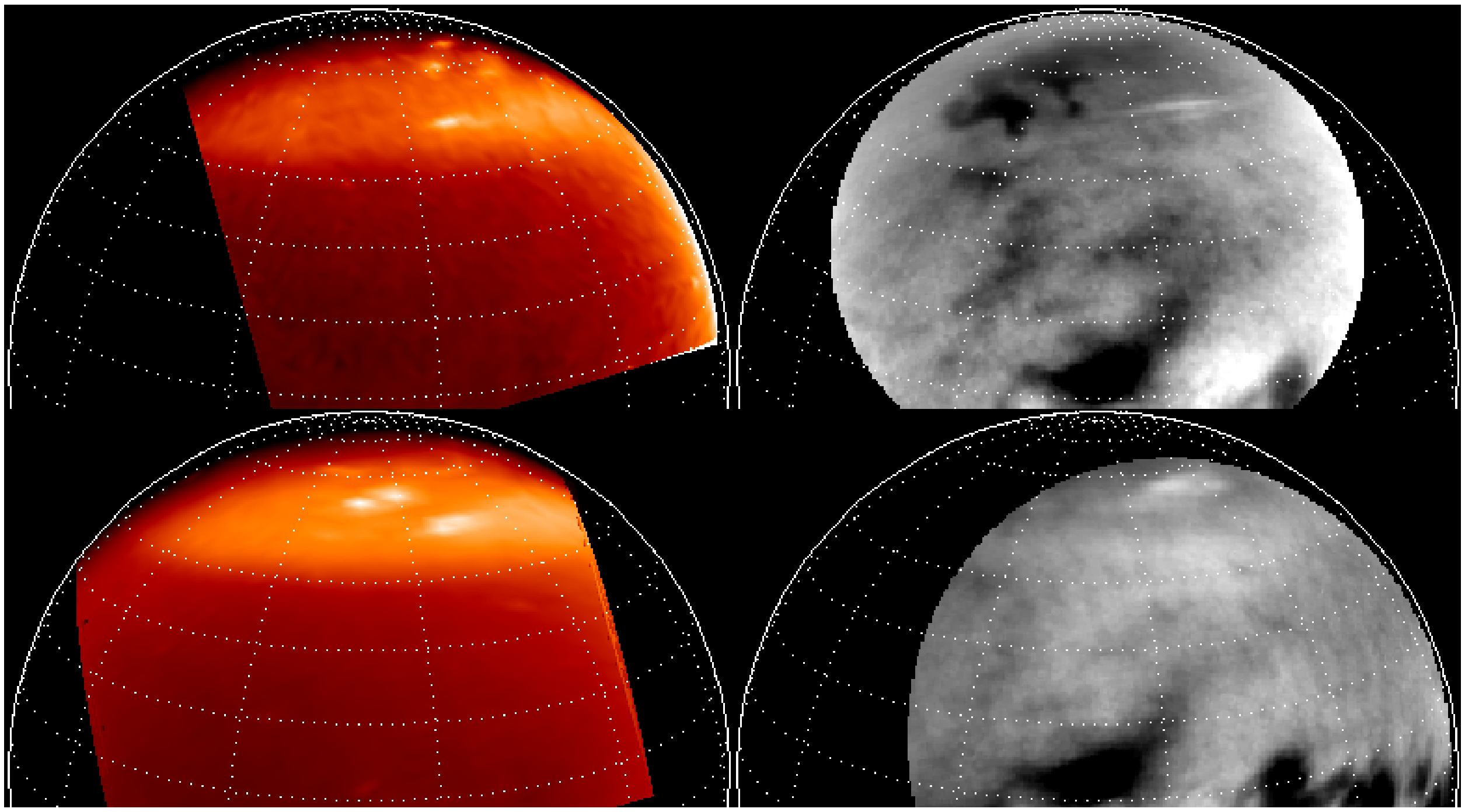

Figure 1. Map projected images of lake-effect clouds at the winter north pole of Titan from the VIMS (left, both from 27 April 2007) and ISS (right, from 24 Feb 2007, top, and 13 April 2007, bottom) imagers on board the Cassini spacecraft. A latitude-longitude grid is imposed on all images. Lines of latitude are shown every 10 degrees, with the north pole barely visible at the top. Lines of longitude are shown every 30 degree; the VIMS images are projected with a central meridian longitude of 140 E, while the ISS images are projected with a central meridian longitude of 80 E. The VIMS images are a sum of 5 images between 2.83 and 2.90 microns, which are wavelengths that exclusively probe regions of Titan at the troposphere and higher. These images show the general north polar cloud previously observed [*Griffith et al.*, 2006] as a general brightening northward of 50 degrees latitude. The lake-effect clouds are visible as spatially and temporally variable brightening within this region. The two VIMS images were obtained with 5.5 hours of each other. Lake-effect clouds are more difficult to discern in the ISS data because of the lack of appropriate filters to exclusively probe higher levels in the atmosphere, nonetheless these clouds can be detected through their morphology, brightness, and variability.

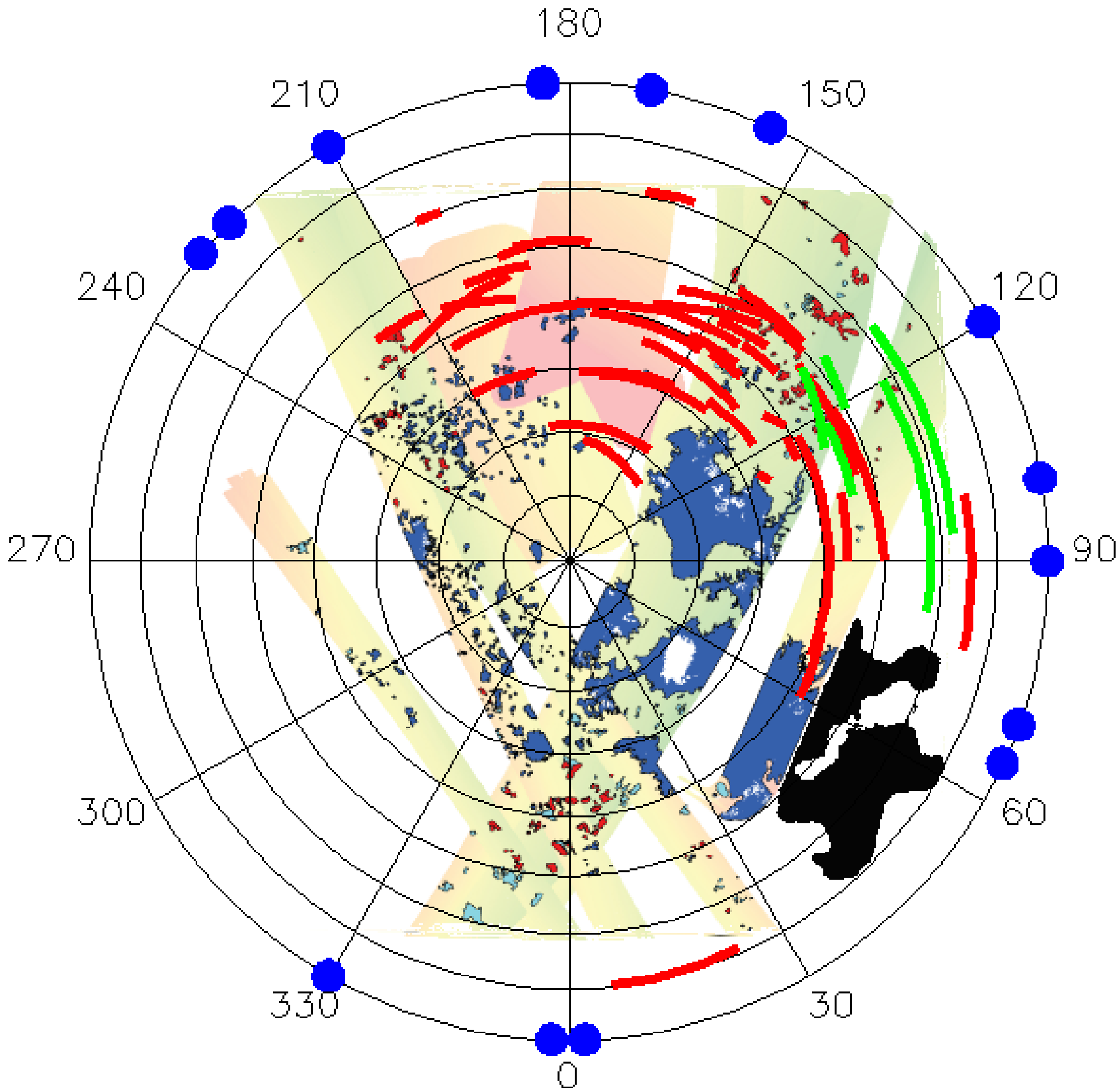

Fig 2. Locations of detected clouds plotted on a polar projection of radar detections of lake and lake-like features at the north pole of Titan [*Hayes et al.*, 2008] Lines of latitude are shown every 5 degrees. The southern extent of the large lake visible in the ISS image in Figure 1 is projected in black onto the radar map to examine the southern extent of the large lakes. Detections from VIMS are shown as red lines, while detections from ISS are shown as green lines. The blue dots at 50 degrees latitude show the approximate longitude of the detections of north polar brightening from ground-based Gemini adaptive-optics images. With the low resolution and poor viewing geometry of the ground-based images, no accurate latitude can be measured.

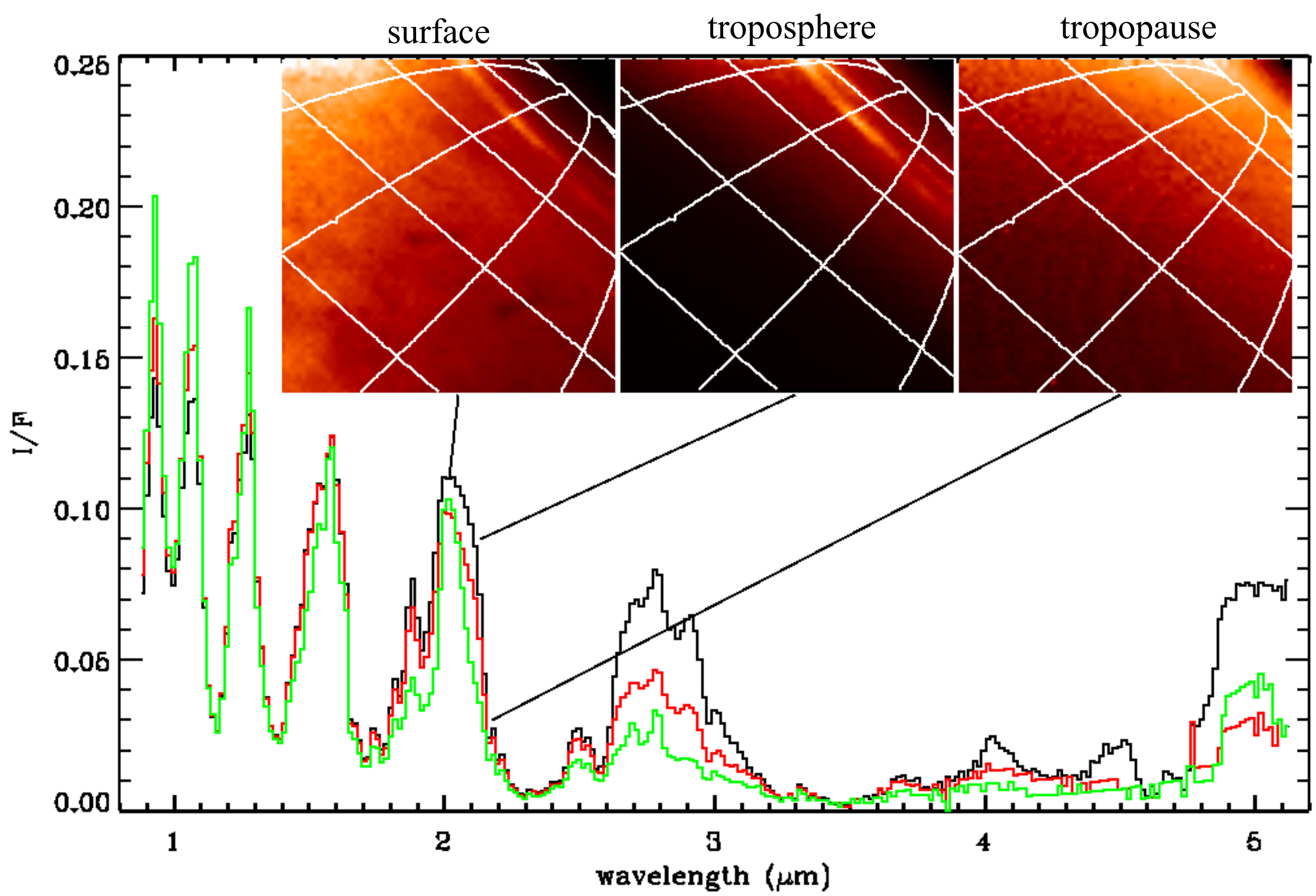

Figure 3. VIMS images and spectra of a fortuitous high spatial resolution observation of a north polar streak. The images show three wavelengths of the hyperspectral images which are sensitive to three different locations in the atmosphere. Latitude lines between 40 and 80 N degrees are shown at 10 degrees intervals, and longitude lines between 130 and 150 E are shown at 10 degree intervals. At 2.02 microns little methane opacity exists so the image shows features all the way to the surface. At 2.13 microns, in a region of moderate methane opacity, the surface is no longer visible but the streaks and north polar cloud are clearly seen. At a wavelength of 2.17 microns, where methane opacity is high, the streaks can no longer be seen, but the north polar cloud is visible. At 2.3 microns (not shown) the methane opacity is so high that no features can be seen. The spectra show isolated regions of the image. The black line shows the spectrum of a bright streak, the green line shows the spectrum of the north polar cloud outside of the streak, while the red line shows the spectrum of the region below 50 degrees N latitude where no cloud is present. The streaks are significantly more reflective at 3 and 5 microns than the surface or north polar cloud, while the north polar cloud is brighter at wavelengths shorter than 2 microns.